\documentclass[a4paper,twocolumn,superscriptaddress,11pt,accepted=2020-06-04]{quantumarticle}
\pdfoutput=1
\usepackage[numbers,sort&compress]{natbib}
\usepackage{epsfig}
\usepackage{amsfonts}
\usepackage{amsmath}
\usepackage{amssymb}
\usepackage{amsthm}
\usepackage{color}
\usepackage{multirow}
\usepackage[normalem]{ulem}
\newcommand{\stkout}[1]{\ifmmode\text{\sout{\ensuremath{#1}}}\else\sout{#1}\fi}
\usepackage{latexsym}
\usepackage{mathrsfs}
\usepackage{natbib}
\usepackage{verbatim}
\usepackage[T1]{fontenc}
\usepackage{float}
\usepackage{rotating}

\usepackage{graphicx}
\usepackage{xcolor}

\usepackage[colorlinks=true,linkcolor=blue,citecolor=magenta,urlcolor=blue]{hyperref}

\DeclareMathOperator{\Tr}{tr}

\newcommand{\ket}[1]{|#1\rangle}

\newcommand{\braket}[2]{\langle#1|#2\rangle}
\newcommand{\bracket}[3]{\langle#1|#2|#3\rangle}
\newcommand{\ketbra}[2]{|#1\rangle\langle#2|}

\begin{document}
	

	\title{The Platonic solids and fundamental tests of quantum mechanics}
	
	
	\author{Armin Tavakoli}
	\affiliation{D\'epartement de Physique Appliqu\'ee, Universit\'e de Gen\`eve, CH-1211 Gen\`eve, Switzerland}

	\author{Nicolas Gisin}
	\affiliation{D\'epartement de Physique Appliqu\'ee, Universit\'e de Gen\`eve, CH-1211 Gen\`eve, Switzerland}

	\begin{abstract}
		The Platonic solids is the name traditionally given to the five regular convex polyhedra, namely the tetrahedron, the octahedron, the cube, the icosahedron and the dodecahedron. Perhaps strongly boosted by the towering historical influence of their namesake, these beautiful solids have, in well over two millennia, transcended traditional boundaries and entered the stage in a range of disciplines. Examples include  natural philosophy and mathematics from classical antiquity, scientific modeling during the days of the European scientific revolution and visual arts ranging from the renaissance to modernity. Motivated by mathematical beauty and a rich history, we consider the Platonic solids in the context of modern quantum mechanics. Specifically, we construct Bell inequalities whose maximal violations are achieved with measurements pointing to the vertices of the Platonic solids. These Platonic Bell inequalities are constructed only by  inspecting the visible symmetries of the Platonic solids. We also construct Bell inequalities for more general polyhedra and find a Bell inequality that is more robust to noise than the celebrated Clauser-Horne-Shimony-Holt Bell inequality.  Finally, we elaborate on the tension between mathematical beauty, which was our initial motivation, and experimental friendliness, which is necessary in all empirical sciences.
	\end{abstract}
	
	
	\maketitle
	

	\section{Introduction}
\begin{figure}
		\centering
		\includegraphics[width=\columnwidth]{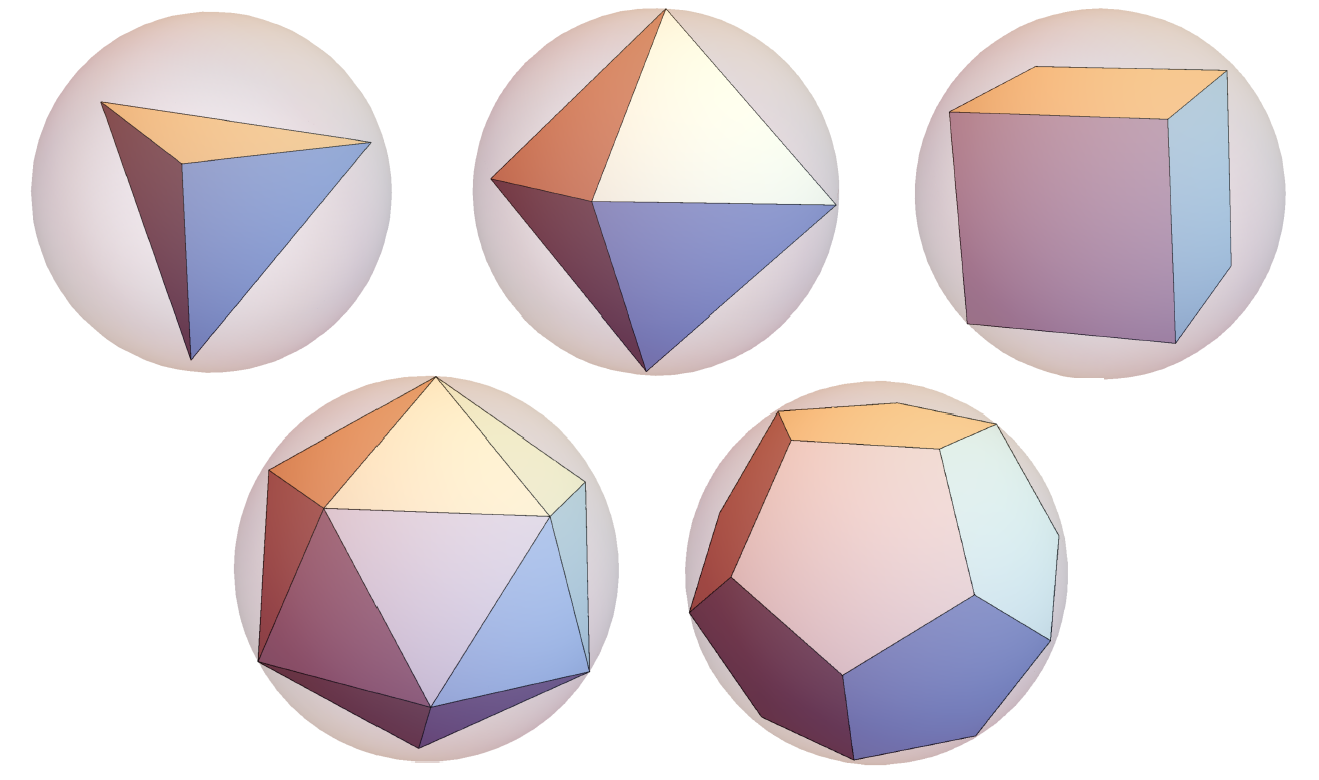}
		\caption{The five Platonic solids inscribed in spheres. From left to right: the tetrahedron, the octahedron, the cube, the icosahedron and the dodecahedron.}\label{AllPlatonicSolids}
	\end{figure}

	Which physicist has never been attracted by mathematical beauty? And what is more beautiful than the Platonic solids; the five regular polyhedra in our three-dimensional space (see Fig.\ref{AllPlatonicSolids})? Here, we first present the fascinating history of these solids and then use them to derive simple Bell inequalities tailored to be maximally violated for measurement settings pointing towards the vertices of the Platonic solids. In this way, we connect beautiful mathematics with foundational quantum physics. However, these \textit{Platonic Bell inequalities} do not distinguish themselves with regard to experimental friendliness: quantum theory predicts that their violations are less robust to noise than the much simpler Clauser-Horne-Shimony-Holt (CHSH) Bell inequality \cite{CHSH}. In fact, Platonic Bell inequalities require more measurement settings - as many as the number of vertices of the platonic solid - than the CHSH Bell inequality, which requires only the absolute minimum of two settings per side. We also construct Bell inequalities tailored to another class of elegant polyhedra, namely the	 Archimedean solids, i.e.~the semi-regular polyhedra. In particular we consider the famous Buckyball, a polyhedron which corresponds to the carbon-60 molecule used in the first molecular interferometer \cite{Arndt} 
	, which requires even more measurement settings. However, we  find that these Bell inequalities also do not offer notable experimental advantages. Finally, we depart from Bell inequalities motivated by mathematical beauty and instead focus our research on finding experimentally friendly Bell inequalities: starting from the Buckyball we iteratively search for noise robust  Bell inequalities. This leads us to a Bell inequality that is somewhat more noise tolerant than the CHSH Bell inequality. However, it is remarkably inelegant. We conclude with a discussion of the danger for theoretical physics to become - and remain - too focused on mathematical beauty \cite{Hossenfelder} at the expense of developing connections with experiments.
	
	\section{A brief history of the Platonic solids in arts, philosophy and science}
	This section provides a broader context  for the Platonic solids. Readers interested exclusively in Bell inequalities may jump to the next section.
	
	The ancient Greek civilisations laid the foundations of western natural philosophy. The development of the latter is permeated by a fascination for geometry. The magnum opus of Greek geometry, Euclid's \textit{Elements}, remained a standard textbook until the 20th century \cite{Wilson}. First printed in Venice in 1482 as one of the earliest mathematics books set in type, it has since been re-printed in at least a thousand editions\footnote{Ref.~\cite{Boyer}, authored in 1968, suggests that the Elements is only outdone in number of editions by the Bible.} and is certainly the most influential mathematical work in history \cite{Boyer}. Geometry allowed the early natural philosophers to describe, understand and make predictions about, the physical world. In the sixth century BC, Thales of Miletus, often hailed as the first scientific philosopher in western civilisation, likely used his knowledge of geometry to measure the height of the pyramids of Egypt \cite{Proklos}. Centuries later, in the Hellenistic period, Eratosthenes accurately calculated  the circumference of the Earth and Hipparcus discovered the precession of the equator. Archimedes' geometry led him to the Law of the Lever \cite{Archimedes}, still taught to every pupil in physics class.

	Geometry was often ascribed a deeper meaning, beyond pure mathematics and its applications. This entails attributing  spiritual, religious or philosophical meaning to certain proportions, planar shapes and solids, elevating the geometries to a tangibly sacred status. The perhaps most famous example of such metaphysical beliefs is due to the Pythagoreans\footnote{For instance, the number three was an ideal number as it was the number of vertices in a triangle, which was a symbol of Apollo. The number ten was termed a perfect number due to the number of vertices in a geometry called a tetractys. The number was therefore honoured by the Pythagoreans not gathering in groups of more than ten people.} \cite{Pythagoras}. Their ideas of sacred geometries were  influential, notably also on key figures such as Plato in the fifth century BC.  In \textit{The Republic}, Plato writes that \textit{"geometry will draw the soul towards truth, and create the spirit of philosophy"} \cite{TheRepublic}. In \textit{Timaeus}, Plato makes concrete the link between geometry and natural philosophy; he discusses the five regular polyhedra, i.e.~the polyhedra whose vertices are identical and whose faces are identical regular polygons, namely the tetrahedron, the octahedron, the cube, the icosahedron and the dodecahedron. Today, these five solids are known as the \textit{Platonic solids} (see Fig.~\ref{AllPlatonicSolids}). Plato assigned four of the solids to the four classical elements thought to be the fundamental form of all matter; the tetrahedron to fire, the octahedron to air, the cube to earth and the icosahedron to water. To the remaining fifth solid, Plato left the following mysterious comment \cite{Timaeus} \textit{"A fifth regular solid still exists, namely the dodecahedron, which does not form the element of any substance; but God used it as a pattern for dividing the zodiac into its twelve signs."} Later, his pupil Aristotle added a fifth element to the original four elements, namely the aether\footnote{Aether theories persisted in science until the strong negative evidence put forward by the Michelson-Morley experiment, performed in 1887.}. It historically became associated to the dodecahedron, perhaps due to its relevance for the golden ratio. From a purely mathematical standpoint, the Platonic solids were the focus of the 13'th book of Euclid's \textit{Elements} which studies their construction and their proportions when inscribed in a sphere.

	The Platonic solids can be appreciated by modern mathematicians for their appealing geometric properties, by modern natural scientists for their occurrence in nature, historical scientific models and metaphysical ideas, and by a broader modern audience for their historical appearance in western visual arts and natural philosophy, as well as their sheer beauty. It appears reasonable to say that the historical interest in the Platonic solids was substantially aided by the fact they were so strongly endorsed by a character as titanic as Plato.
	
	Almost two millennia after Plato, the maintained appreciation for the Platonic solids could for instance be seen in Luca Pacioli's mathematics book \textit{De Divina Proportione}. Published in 1509, it spends its first section motivating the divinity of the golden ratio; in particular by emphasising that the golden ratio appears in the dodecahedron, which is a representation of the aether \cite{Livio}. The book's lasting success even outside mathematics circles may in part be due to its masterful illustrations of the Platonic solids and various other geometries, in drawings signed Leonardo da Vinci. In fact, the works of many artists feature the Platonic solids; ranging from the renaissance mosaics in the cathedral of San Marco in Venice to the 20th century works of Maurits Escher, who incidentally also kept a coveted model of the nested Platonic solids in his office \cite{MagicMirror}. Salvador Dal\'i's 1955  painting \textit{The Sacrament of the Last Supper} (framed in the golden ratio) sets stage  inside a dodecahedron.
	
	In the realm of natural philosophy, the Platonic solids found a new role in the 1597 publication of \textit{Mysterium Cosmographicum} authored by Johannes Kepler. Kepler proposed a model of the heliocentric solar system in which the six known planets were modeled by nesting the five Platonic solids and inscribing and circumscribing them by spheres \cite{Kepler}.  Although this model was later abandoned due to its inconsistencies with astronomical observations, it served as a stepping stone to Kepler's three laws of planetary motion. Albeit not in the solar system, the Platonic solids present themselves elsewhere in nature. Three of them are natural structures of crystals. A range of Boron compounds include Boron-12 which takes an icosahedral form. The icosahedron is also the structure of many species of Radiolaria and viruses, e.g.~polio. Curiously, it was the discovery of the icosahedral phase in quasicrystals  that led to the Nobel prize in chemistry in 2011 \cite{QuasiCrystal}. Notably, the most common silicates are structured as a silicon atom binding to four oxygen atoms. The silicon atom sits at the center of a tetrahedron with the oxygen atoms sitting at its vertices. Interestingly, silicates comprise the majority of Earth's crust and mantle, and they are often the  dominating mineral in various forms of soil. Perhaps, had Plato ascribed the tetrahedron rather than the cube as the manifestation of earth, his metaphysical ideas might have better withstood the test of time.

	\section{A brief history of Bell inequalities}
	This section provides a non-technical introduction to Bell inequalities. Readers interested mainly in the technical considerations may proceed immediately to the next section. 
	
	Modern science, with its emphasis on empiricism, has for long left behind ideas of Euclidean geometry being fundamental to describing nature. The 19th century saw the development of curved (non-euclidean) geometry\footnote{Non-euclidean geometry was the climax of two millennia of mathematical discussions, first led by Greeks, then by Arabs and Persians and finally by renaissance Europeans, about Euclid's fifth postulate (parallel lines) \cite{Noneuclid}.} which in the early 20th century found a fundamental role in  Einstein's theory of gravity. The 20th century also brought with it the perhaps most radical change of scientific paradigm since the days of Newton, namely the theory of quantum mechanics, which governs nature on the scale of atoms and elementary particles. The most radical predictions of quantum mechanics defied the \textit{principle of locality}, i.e.~that events that are very far separated in space and time cannot influence each other\footnote{It is interesting to point out that some earlier theories such as Newtonian gravity in fact did not respect the principle of locality; gravity propagates instantaneously. This was, however, generally perceived as a major drawback.} \cite{Chance}. This counterintuitive feature put quantum mechanics on an apparent collision course with the famous no-signaling principle.
	
	Quantum mechanics can be understood to  claim that two objects, separated by large distances could still influence each other. Take a pair of atoms, which have a magnetic moment due the angular momentum and spin of their electrons and nucleus. We measure the direction of the atom's magnetic moment. Quantum mechanics tells us that if we were to find the magnetic moment of the first atom pointing upwards, then this can change the magnetic moment of our second atom so that it will also be found pointing upwards. This influence is \textit{immediate}, and does not even require some carrier (e.g.~a mechanical wave or light) to bring it from one atom to the other. Today, this phenomenon is famous under the name \textit{entanglement}: the fact that the whole system is greater than the collection of its individual parts. Remarkably, however, quantum mechanics still manages to peacefully coexist with the principle of no-faster-than-light communication. The reason is that although distant systems influence each other,  the influence does not carry any information from one system to the other. In the 1920s, the question of whether entanglement exists prompted an intensive series of debates between Einstein and Bohr; the former speaking of a ``spooky action at a distance``, and the latter in support of quantum mechanics. 
	
\begin{figure}
		\centering
		\includegraphics[width=\columnwidth]{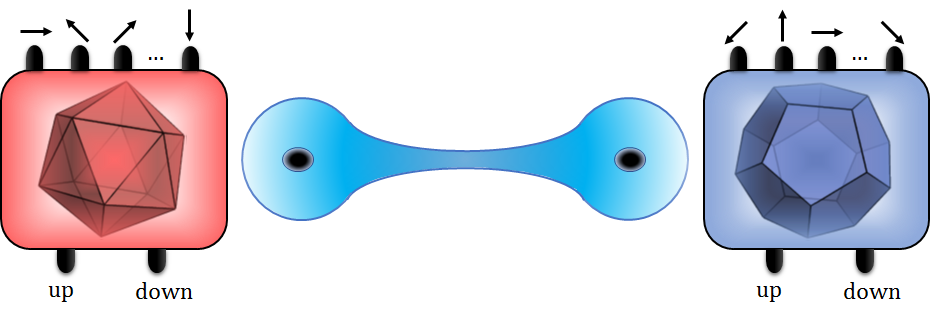}
		\caption{Illustration of a Bell experiment. Two separate atoms that are entangled with each other are sent to different stations where their magnetic moments are measured along various directions. Each measurement answers whether the magnetic moment points up or down the axis along which it is measured. In a Platonic Bell inequality, the best measurements at each station are those that form a Platonic solid.}\label{FigBell}
	\end{figure}

	Nevertheless, and most remarkably, in 1964 physicist John Bell  proved that the existence of entanglement could in fact be scientifically settled \cite{Bell}. Bell found a way of capturing the essence of what local theories predicted about the correlations between the magnetic moments. For example, if one finds the first magnetic moment pointing in some direction,  how often does one also find the second magnetic moment pointing in the same direction? If the former points to the left, to what extent does it mean that the latter will be pointing right? Answering such questions tells us the \textit{correlations} between the two distant magnetic moments. Bell showed that some correlations that were possible in quantum mechanics were in fact impossible in local theories; local correlations obey relations today known as \textit{Bell inequalities}, which can be violated in quantum theory \cite{RMP}. The existence of entanglement could therefore be confirmed by an experiment (see Fig.~\ref{FigBell} for an illustration of a Bell experiment) successfully violating a Bell inequality. Early experiments strongly supported quantum mechanics \cite{Clauser, Aspect} and the matter was definitely settled by experiments in 2015 \cite{LoopholeFree}. The monumental violation of Bell inequalities established entanglement as a natural phenomenon which gave rise to the today rapidly developing field of quantum information theory. This field promises things such as quantum computers, quantum cryptography and teleportation as exciting technologies in a currently unraveling ``second quantum revolution`` \cite{AspectRevolution}.

	\section{The Platonic solids}
	A three-dimensional solid that has sharp corners, straight edges and polygonal faces is called a polyhedron. The Platonic solids is the umbrella term for all polyhedra that are both convex and regular. In an intuitive but informal way, these terms mean the following:
	\begin{itemize}
		\item \textbf{Convex polyhedron:} every two points inside the polyhedron can be connected with a straight line that itself is inside the polyhedron.
		\item \textbf{Regular polyhedron:} the edges, vertices and faces respectively look the same.
	\end{itemize}
	In two dimensions, it is easily seen that there are infinitely many regular convex polygons. Remarkably, the situation changes completely in three dimensions; Euclid proved that there are only five regular convex polyhedra. These are called the Platonic solids (see Fig.~\ref{AllPlatonicSolids}). Let us briefly review each of them. 
	\begin{itemize}
		\item \textbf{Tetrahedron.} A triangular pyramid with four faces, four vertices and six edges.
		\item \textbf{Octahedron.} 	A triangular antiprism with eight faces, six vertices and twelve edges.
		\item \textbf{Cube.} A box with six square faces, eight vertices and twelve edges.
		\item \textbf{Icosahedron.} 20 triangular faces, twelve vertices and 30 edges. By dividing its  vertices suitably in three sets of four, one can inscribe three perpendicular golden rectangles.
		\item \textbf{Dodecahedron.} Twelve pentagonal faces, 20 vertices and 30 edges. Its surface area, volume and distance between adjacent vertices are related to the golden ratio. 
	\end{itemize}
	
	To every polyhedron, we can associate a partner polyhedron called its dual. The dual of the dual is again the original polyhedron. To construct the dual of a polyhedron, the main idea is to let the vertices of the dual pass through the midpoint of the faces of the original polyhedron. The Platonic solids exhibit particularly elegant duality relations: the tetrahedron is its own dual whereas the octahedron and cube are dual to each other and similarly for the icosahedron and  the dodecahedron. Thus, the dual of a Platonic solid is always a Platonic solid.

	\section{Bell inequalities}
	The magnetic moment of an atom is a direction in three-dimensional space; we can think of it as an arrow denoted $\vec{n}$ on a unit-radius sphere. Imagine that  we want to measure the magnetic moment. This can be done along any axis we want, labeled by an arrow $\vec{m}$ on our sphere. Quantum mechanics tells us how to compute the probability of our magnetic moment, initially in direction $\vec{n}$, being found up (along the positive axis) and down (along the negative axis) respectively,  when measured along $\vec{m}$.
	
	Let us now add a second atom. We separate the pair, sending one atom to Alice and one atom to Bob. Alice may measure the magnetic moment of her atom in various directions. Let us say that she has $N_\text{A}$ different directions to choose from. We label her choice of measurement direction $x=1,\ldots,N_\text{A}$ and label the corresponding direction by $\vec{a}_x$. Similarly, Bob may measure his magnetic moment in one of $N_\text{B}$ different directions. We label his choice of direction $y=1,\ldots,N_\text{B}$ and the specific direction by $\vec{b}_y$. For given choices of measurements, there are four possible outcomes. These are $++,+-,-+$ and $--$. If Alice and Bob have the same outcome, i.e.~either $++$ or $--$, we say that Alice and Bob are correlated. If they have different outcomes, either $+-$ or $-+$, we say that Alice and Bob are anticorrelated. It is therefore handy to introduce a correlator which captures the degree of correlation or anticorrelation; 
	\begin{equation}
	E(x,y)=p(+,+)+p(-,-)-p(+,-)-p(-,+).
	\end{equation}
	The closer $E$ is to one (negative one), the stronger are the correlations (anticorrelations). When $E=0$ there are no correlations between the outcomes.

	We wish to determine whether the correlations contained in the list $\{E(x,y)\}_{x,y}$  can be explained by local theories. To this end, we must construct Bell inequalities. These are inequalities of the form  
	\begin{equation}\label{Bell}
	\mathcal{B}\equiv \sum_{x=1}^{N_\text{A}}\sum_{y=1}^{N_\text{B}}c_{x,y}E(x,y)\stackrel{\text{local}}{\leq} C,
	\end{equation}
	where $c_{x,y}$ are some real numbers and $C$ is a bound that is respected by \textit{all possible local theories}. We emphasise that the local bound holds irrespective of the measurement directions used to obtain the expectation values.
	
	What does it mean that the correlations can be modeled with a local theory? Local models assume that when the particles were created, they were endowed with some shared property $\lambda$. A measurement simply reveals that already existing property. If Alice chooses measurement $x$, a local model determines whether the outcome is $+$ or $-$ given the property $\lambda$. The analogous goes for Bob. However, we do not know what specific property $\lambda$ represents. Our ignorance of it is represented by a probability distribution $p(\lambda)$. Therefore, in a local model, the correlators reads
	\begin{equation}\label{lhv}
	E(x,y)=\sum_{\lambda} p(\lambda) E^\text{A}_\lambda(x)E^\text{B}_\lambda(y).
	\end{equation}
	Thus, to find the local bound $C$  in Eq.~\eqref{Bell}, we must maximise $\mathcal{B}$ over $p(\lambda)$. Fortunately, this can be determined by checking a finite number of specific choices of $p(\lambda)$ (all the deterministic responses of Alice and Bob) and pick the largest one \cite{Fine}. 
	
	The critical point is that Bell inequalities can sometimes be violated ($\mathcal{B}>C$) if the Bell experiment is modeled within quantum mechanics, i.e.~by Alice and Bob having their two magnetic moments in an entangled state. The most interesting case is when the two magnetic moments are \textit{maximally entangled}, i.e.~in the state
	\begin{equation}
	\ket{\phi^+}=\frac{\ket{\uparrow\uparrow}+\ket{\downarrow\downarrow}}{\sqrt{2}}.
	\end{equation}
	This state has the remarkable property that if Alice measures her magnetic moment along direction $\vec{n}$, the magnetic moment of Bob ends up also pointing either up or down the axis $\vec{n}$ (up to a reflection in the $xz$-plane). This paves the way for quantum correlations that violate the Bell inequality and therefore do not admit a local model. The natural question becomes, how strong can quantum correlations be? How much can they violate a Bell inequality? In what follows, we construct Bell inequalities that achieve their maximal correlations in quantum mechanics by Alice and Bob choosing their measurement directions $\vec{a}_x$ and $\vec{b}_y$ to respectively point to the vertices of a Platonic solid. 
	
	By \textit{Platonic Bell inequality}, we mean to say a Bell inequality that is maximally violated in quantum theory with measurements forming pairs of Platonic solids (see Fig.~\ref{FigBell}). Notably, Platonic solids have previously been used in the context of quantum mechanics, e.g.~to construct correlation tests for a phenomenon known as steering \cite{Steering}, which is a weaker notion of a genuinely quantum phenomenon, as compared to the violation of Bell inequalities.

	\section{Two simple Platonic Bell inequalities}
	We begin by presenting two particularly simple Platonic Bell inequalities. Their simplicity stems from the fact that all the coefficients $c_{x,y}$ appearing in Eq.~\eqref{Bell} are either $+1$, $-1$ or $0$, and that the Bell inequalities are constructed by inspecting the symmetries between a Platonic solid and its dual Platonic solid. Our first Platonic Bell inequality gives Alice and Bob measurement settings that correspond to a cube and an octahedron respectively (being dual polyhedra). Our second Platonic Bell inequality is based on the icosahedron and the dodecahedron (again being dual polyhedra).

	\subsection{The first Platonic Bell inequality}
\begin{figure}[t]
		\centering
		\includegraphics[width=0.75\columnwidth]{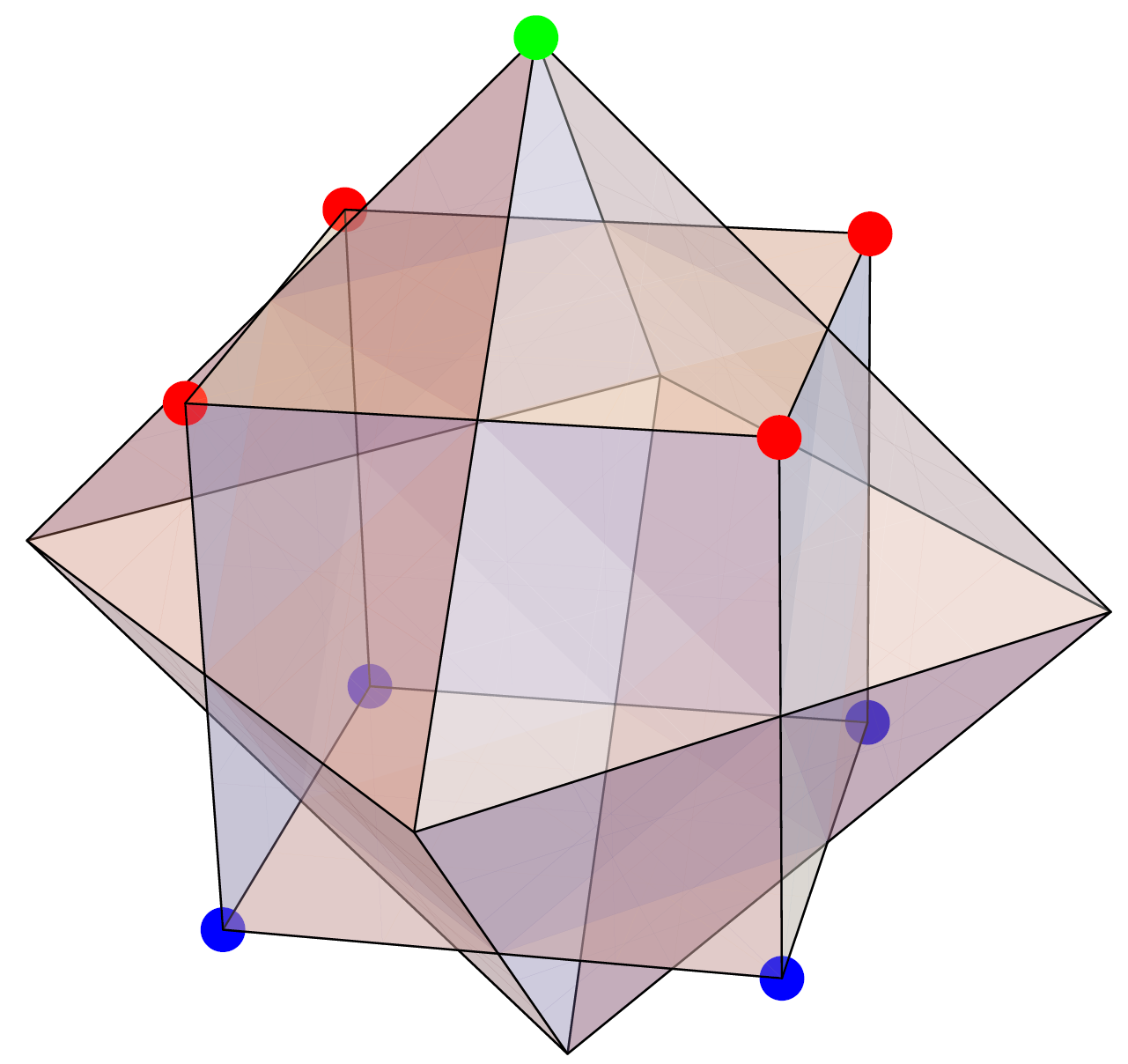}
		\caption{A compound of two dual Platonic solids: the cube and the octahedron. For each vertex of the octahedron (for example green point), four vertices of the cube are equally close to it (red points) whereas the remaining four vertices of the cube are equally distant to it (blue points).}\label{firstPlatonic}
	\end{figure}

	We construct a Platonic Bell inequality for the cube and the octahedron. To this end, we consider a Bell experiment in which Alice has eight possible settings which we label by a three-bit string $x=x_1x_2x_3\in\{0,1\}^3$ and Bob has six possible settings $y=y_1y_2$ which we label by a trit $y_1\in\{1,2,3\}$ and a bit $y_2\in\{0,1\}$. In order to construct the Bell inequality, we visualise a compound of a cube and an octahedron (see Fig.~\ref{firstPlatonic}). The fact that these solids are dual to each other makes the compound highly symmetric. We exploit this to construct our Platonic Bell inequlity.

	We now reason as follows. If Alice's and Bob's magnetic moments are maximally entangled, it means that if Alice measures her magnetic moment in the direction corresponding to the vertex of the octahedron (green point) and finds the outcome (say) $+$, she will remotely prepare Bob's magnetic moment in the same state (up to reflection in the $xz$-plane) as that into which her state has collapsed. Also, if a magnetic moment points in direction $\vec{n}$ and is measured along $\vec{m}$, the correlations (anticorrelations) are stronger the closer (more distant) the two vectors are. Four of the vertices of the cube (red points) are close, and equally close, to the vertex of the octahedron (green point). Therefore, we let the reasonably strong correlations contribute towards our Bell inequality test; specifically we put  $c_{x,y}=1$. Similarly, the other four vertices of the cube (blue points) are distant, and equally distant, from the vertex of the octahedron (green point). Hence, we let the reasonably strong anticorrelations contribute towards our Bell test; we put $c_{x,y}=-1$. Repeating this reasoning for every vertex of the octahedron, we arrive at the first Platonic Bell inequality. It reads 
	\begin{equation}\label{firstineq}
	\mathcal{B}_\text{cuboct}=\sum_{x,y} (-1)^{x_{y_1}+y_2}E(x,y)\stackrel{\text{local}}{\leq} 24.
	\end{equation}
	The local bound is obtained by considering all assignments of $+$ and $-$ to the outcomes of Alice and Bob. To derive it, we write $A_x,B_y\in\{\pm 1\}$ and impose the form of Eq.~\eqref{lhv}. This gives
	\begin{multline}
	\mathcal{B}_\text{cuboct}=\\
	\sum_{x}A_x \sum_y (-1)^{x_{y_1}+y_2}B_y\leq \sum_x\left|\sum_y (-1)^{x_{y_1}+y_2}B_y\right|\\
	=\sum_x\Big|(-1)^{x_1}(B_{10}-B_{11})+(-1)^{x_2}(B_{20}-B_{21})\\
	+(-1)^{x_3}(B_{30}-B_{31})\Big|.
	\end{multline}
	Notice that for all $y_1$, we have $B_{y_10}-B_{y_11}\in\{-2,0,2\}$. A little inspection shows that it is optimal to never choose the value zero. In fact, as long as we choose $B_{y_10}-B_{y_11}=\pm2$, we always find the local bound $\mathcal{B}_\text{cuboct}=24$. We remark that the Bell inequality \eqref{firstineq} is closely related to the so-called Elegant Bell inequality \cite{GisinElegant}; the settings of Alice and Bob are merely doubled.

	Now, in order to show that we indeed have a Platonic Bell inequality, we must derive the maximal quantum violation and show that it is achievable with a cube on Alice's side and an octahedron on Bob's side. If we let Alice and Bob share the maximally entangled state and perform measurements corresponding to these Platonic solids, we find that
	\begin{equation}\label{firstQ}
	\mathcal{B}_\text{cuboct}=16\sqrt{3}\approx 27.71,
	\end{equation}
	which is a violation of the Bell inequality.
	
	Let us now prove that no larger value is possible in quantum theory i.e.~there exists no entangled state (of potentially higher dimension) and no local measurements that can generate a larger Bell inequality violation. We write
	\begin{equation}
	\mathcal{B}_\text{cuboct}=\sum_x \braket{\alpha_x}{\beta_y}
	\end{equation}
	where 
	\begin{align}
	& \ket{\alpha_x}=A_x\otimes\openone \ket{\psi}\\
	& \ket{\beta_x}=\openone \otimes \sum_y (-1)^{x_{y_1}+y_2}B_y\ket{\psi}.
	\end{align}
	Here $A_x$ is a general observable of Alice and $B_y$ is a general observable of Bob. We use the Cauchy-Schwarz inequality,  the fact that $\braket{\alpha_x}{\alpha_x}=1$ and a simple concavity inequality to write
	\begin{equation}\label{bound}
	\mathcal{B}_\text{cuboct}\leq \sum_x \sqrt{\braket{\beta_x}{\beta_x}}\leq \sqrt{8}\sqrt{\sum_x \braket{\beta_x}{\beta_x}}.
	\end{equation}
	Let us now consider the sum under the square-root on the right-hand-side. We find
	\begin{align}\nonumber
	&\sum_x \braket{\beta_x}{\beta_x}=\sum_x \sum_{y,y'} (-1)^{x_{y_1}+x_{y_1'}+y_2+y_2'}\bracket{\psi}{B_yB_{y'}}{\psi}\\\nonumber
	&=\sum_{y,y'}(-1)^{y_2+y_2'} \left(\sum_x(-1)^{x_{y_1}+x_{y_1'}} \right)\bracket{\psi}{B_yB_{y'}}{\psi}\\\nonumber
	&=8\sum_{y_1,y_2,y_2'} (-1)^{y_2+y_2'}\bracket{\psi}{B_{y_1y_2}B_{y_1y_2'}}{\psi}\\
	&=48-8\sum_{y_1}\bracket{\psi}{\{B_{y_10},B_{y_11}\}}{\psi}\leq 96,
	\end{align}
	where we have used that $B_y^2=\openone$ and that $\{B_{y_10},B_{y_11}\}\geq -2\openone$. Inserting this into Eq.~\eqref{bound}, we recover the quantum bound $\mathcal{B}_\text{cuboct}\leq 16\sqrt{3}$. We conclude that our inequality \eqref{firstineq} indeed is a Platonic Bell inequality for the cube and the octahedron.

	\begin{figure}
		\centering
		\includegraphics[width=0.75\columnwidth]{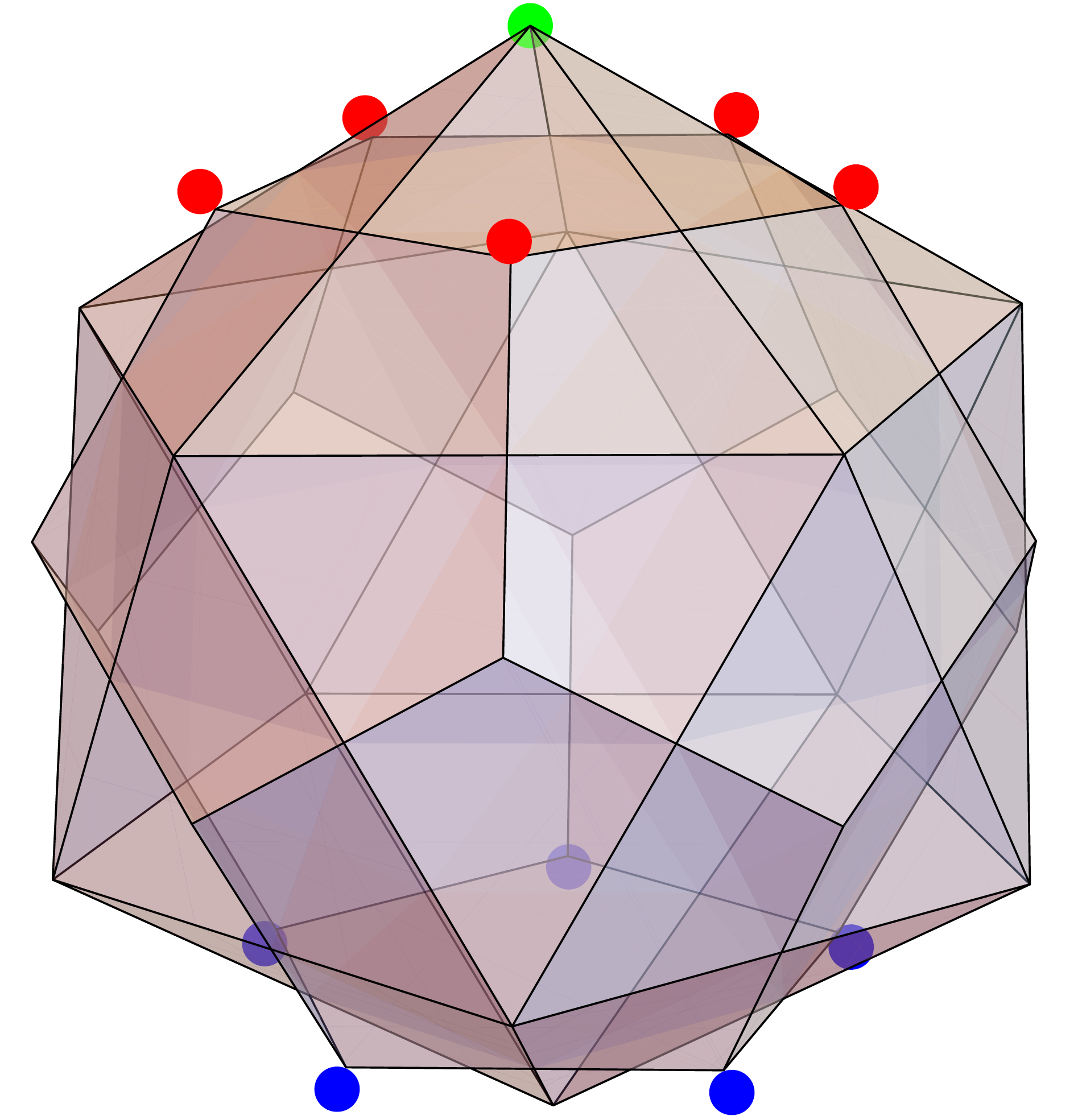}
		\caption{A compound of two dual Platonic solids: the icosahedron and the dodecahedron. For each vertex of the icosahedron (for example green point), five vertices of the dodecahedron are (equally) close to it (red points) whereas another five vertices of the dodecahedron are (equally) distant to it (blue points).}\label{secondPlatonic}
	\end{figure}

	\subsection{The second Platonic inequality}
	Our first Platonic Bell inequality relied on exploiting the duality between the cube and the octahedron. The same intuition can be used to construct a simple Platonic Bell inequality for Alice performing measurements forming an icosahedron and Bob performing measurements forming a dodecahedron. Since the vector antipodal to every vector pointing to a vertex of the icosahedron and the dodecahedron respectively also points to a vertex, we can simplify the setting by only supplying Alice and Bob with a number of settings equal to half the number of vertices in the icosahedron and dodecahedron respectively. This means that we consider a Bell inequality test in which Alice has six settings and Bob has ten settings. In analogy with the previous, we visualise a compound of the icosahedron and the dodecahedron, see Fig.~\ref{secondPlatonic}. Duality presents us with a highly symmetric compound which we exploit to construct our Bell inequality. Again, we imagine that the two magnetic moments are maximally entangled and that Alice therefore remotely prepares Bob's magnetic moment in the same direction as her own once she has measured it. Then, both the magnetic moments will (for example) point to the vertex (green point) of the icosahedron. Since this vertex is close, and equally close, to five vertices of the dodecahedron (red points) while distant, and equally distant, to five other vertices of the dodecahedron (blue points), we reward (put $c_{x,y}=1$) correlations in the first five events and analogously reward  (put $c_{x,y}=-1$) anticorrelations in the latter five events. In the event of Bob measuring in a direction corresponding to a vertex of the dodecahedron which is neither among the five close nor the five distant ones, we give no reward ($c_{x,y}=0$). This simple reasoning leads to a list of coefficients which can straightforwardly be rearranged (permutations and global sign flips) to the coefficients 
	\begin{equation}
	\begin{tiny}
	c^\text{icodod}=\left(
	\begin{array}{cccccccccc}
	1 & 1 & 1 & 0 & 1 & 1 & 0 & 0 & 0 & 0 \\
	1 & 1 & 0 & 1 & 0 & 0 & 1 & 1 & 0 & 0 \\
	1 & 0 & 1 & 1 & 0 & 0 & 0 & 0 & 1 & 1 \\
	0 & 1 & 0 & 0 & 1 & 0 & 1 & 0 & -1 & -1 \\
	0 & 0 & 1 & 0 & 0 & 1 & -1 & -1 & 1 & 0 \\
	0 & 0 & 0 & 1 & -1 & -1 & 0 & 1 & 0 & 1 \\
	\end{array}
	\right)
		\end{tiny}
	\end{equation}
	The corresponding Bell inequality becomes 
	\begin{equation}\label{second}
	\mathcal{B}_\text{icodod}= \sum_{x=1}^{6}\sum_{y=1}^{10}c_{x,y}^\text{icodod}E(x,y) \stackrel{\text{local}}{\leq} 20,
	\end{equation}
	where the local bound is obtained by considering all assignments of outcomes $(+,-)$ to Alice and Bob. 
	
	By sharing the maximally entangled state $\ket{\phi^+}$ and Alice performing measurements corresponding to an icosahedron and Bob performing measurements corresponding to a dodecahedron, we obtain the quantum value
	\begin{equation}
	\mathcal{B}_\text{icodod}=2\sqrt{45+60\varphi}\approx 23.84,
	\end{equation}
	where $\varphi=\frac{1+\sqrt{5}}{2}$ is the golden ratio. We have confirmed the optimality of this value (up to machine precision) using the hierarchy of quantum correlations \cite{NPA}.  This shows that Eq.~\eqref{second} indeed is a Platonic Bell inequality. We note that one can attempt a more standard analytical proof of the quantum bound via the method used to derive the optimality of Eq.~\eqref{firstQ}. However, this is significantly more cumbersome due to the increased number of settings.

	\begin{table*}[t]
		\centering
		\resizebox{1.5\columnwidth}{!}{%
			\begin{tabular}{|c|c|c|c|c|c|}
				\hline
				\multicolumn{1}{|l|}{} & Tetrahedron                         & Octahedron                      & Cube                                & Icosahedron & Dodecahedron  \\ \hline
				Tetrahedron            & 16/3 | 16/3 & 7.82 | 8                        & 9.24 | 32/3                         & 14.78 | 16  & 22.82 | 80/3  \\ \hline
				Octahedron             & -                                   & 12 | 12 & 13.86 | 16                          & 21.96 | 24  & 34.40 | 40    \\ \hline
				Cube                   & -                                   & -                               & 64/3 | 64/3 & 29.89 | 32  & 47.51 | 160/3 \\ \hline
				Icosahedron            & -                                   & -                               & -                                   & 41.89 | 48  & 63.57 | 80    \\ \hline
				Dodecahedron           & -                                   & -                               & -                                   & -           & 109.7 | 400/3 \\ \hline
			\end{tabular}
		}
		\caption{Local (left) and quantum (right) bounds for Bell inequalities for all pairs of Platonic solids. In all cases except that of two tetrahedra, two octahedra and two cubes we find a quantum violation. In all cases but these, we have Platonic Bell inequalities.}\label{tabPlato}
	\end{table*}

	\section{A systematic method}
	Let us now outline a more general approach to the construction of Platonic Bell inequalities. Here, we choose a pair of Platonic solids for Alice and Bob and construct a Bell inequality for which the chosen solids are optimal. 
	
	Let the vectors pointing to the vertices of Alice's Platonic solid be denoted $\{\vec{v}_x\}$. Similarly, the vectors $\{\vec{u}_y\}$ denote the vertices of Bob's Platonic solid. For simplicity, we let Alice have the solid with the smaller number of vertices. Consider now the following Bell inequality
	\begin{equation}\label{gen}
	\mathcal{B}_\text{Plato}\equiv \sum_{x=1}^{N_\text{A}}\sum_{y=1}^{N_\text{B}}(\vec{v}_x\cdot \vec{u}_y^*)E(x,y)\stackrel{\text{local}}{\leq} C,
	\end{equation}
	where $\vec{u}^*=(u^1,-u^2,u^3)$.  That is, we reward correlations and anticorrelations between Alice and Bob by an amount corresponding to the scalar product between the vertices of the desired Platonic solids (up to one being reflected in the $xz$-plane). It is worth noting that the Bell inequality depends on the relative angle between the two Platonic solids, which typically also will influence the local bound. The local bound $C$ can straightforwardly be evaluated by considering all output strategies;
	\begin{align}\nonumber\label{cbound}
	C&=\max_{\substack{A_1,\ldots,A_{N_\text{A}}\in\{\pm 1\}^{N_\text{A}}\\B_1,\ldots,B_{N_\text{B}}\in\{\pm 1\}^{N_\text{B}}}} \sum_{y} B_y \sum_x(\vec{v}_x\cdot \vec{u}_y^*)A_x \\
	&= \max_{A_1,\ldots,A_{N_\text{A}}\in\{\pm 1\}^{N_\text{A}}} \sum_y\left|\sum_x(\vec{v}_x\cdot \vec{u}_y^*)A_x\right|.
	\end{align}
	Thus, we find the local bound by considering $2^{N_\text{A}}$ evaluations.

	Let us now evaluate the value of $\mathcal{B}_\text{Plato}$ in a quantum model in which Alice and Bob share the maximally entangled state $\ket{\phi^+}$.  We let Alice's measurements be represented by the  vectors $ \vec{a}_x=\vec{v}_x$ and Bob's measurements be represented by $\vec{b}_y=\vec{u}_y$. We find
	\begin{align}\label{corr}\nonumber
	\mathcal{B}_\text{Plato}&=\sum_{x,y}(\vec{v}_x\cdot \vec{u}_y^*)\bracket{\phi^+}{A_x\otimes B_y}{\phi^+}\\\nonumber
	& =\sum_{x,y}(\vec{v}_x\cdot \vec{u}_y^*)\bracket{\phi^+}{\openone\otimes B_yA^\text{T}_x}{\phi^+}\\\nonumber
	& =\sum_{x,y}(\vec{v}_x\cdot \vec{u}_y^*)\bracket{\phi^+}{\openone\otimes (\vec{u}_y^*\cdot \vec{\sigma}^\text{T})(\vec{v}_x\cdot\vec{\sigma}^\text{T})}{\phi^+}\\
	& =\sum_{x,y}(\vec{v}_x\cdot \vec{u}_y^*)^2
	\end{align} 
	In the second line, we have used that for any observable $R\otimes\openone\ket{\phi^+}=\openone\otimes R^\text{T}\ket{\phi^+}$ and in the penultimate line we have used that $\Tr\left((\vec{u}_y^*\cdot \vec{\sigma}^\text{T})(\vec{v}_x\cdot\vec{\sigma}^\text{T})\right)=2\vec{v}_x\cdot\vec{u}_y^*$.

	Let us now consider the maximal quantum correlations. We note that there are 15 possible pairs of Platonic solids (including when both solids are the same). For each of these 15 cases, we have constructed the Bell inequality \eqref{gen}, computed the quantum value \eqref{corr} and compared it to the maximal quantum value obtained via the first level of the hierarchy of quantum correlations. We find that the quantum strategy based on the Platonic solids always is optimal. In Table~\ref{tabPlato} we compare the maximal quantum correlations with the local bound. We see that in all cases except for that of two tetrahedra, two octahedra and two cubes, the quantum correlations violate the local bound\footnote{Since the relative angle between the two Platonic solids matters, we specify that the vertices of the Platonic solids where chosen to be the ones given by the software Mathematica's built-in function "PolyhedronData".}\footnote{We remark that the visibility required for a violation in the presence of white noise is the ratio between the local and quantum bounds.}. Moreover, due to the structure of the Platonic Bell inequalities, the maximal quantum value of $\mathcal{B}_{\text{Plato}}$ is a simple rational number.

	\subsection{A Buckyball Bell inequality} 
	\begin{figure}
		\centering
		\includegraphics[width=0.65\columnwidth]{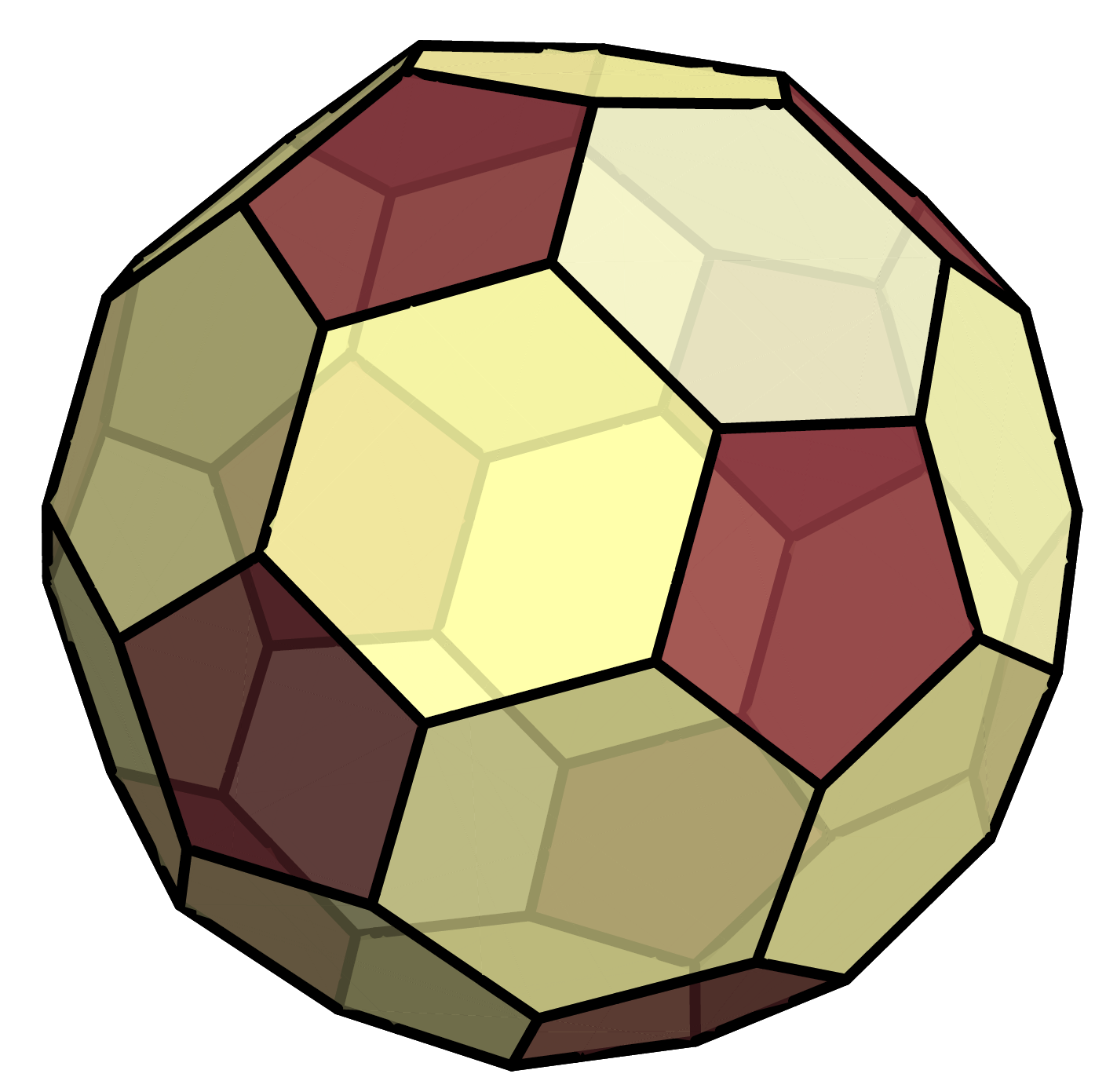}
		\caption{The truncated icosahedron is an Archimedean solid with 32 faces, 60 vertices and 90 edges.}\label{truncatedIcosahedron}
	\end{figure}
	
	The Bell inequality construction \eqref{gen} also works for some polyhedra that are not Platonic solids. Here, we illustrate this fact by considering a so-called Archimedean solid\footnote{The Archimedean solids are the semi-regular convex polyhedra (excluding the Platonic solids, prisms and antiprisms) of which there are 13.}. Specifically, we focus on the solid obtained from cutting an icosahedron symmetrically at every vertex so that each of them is replaced with a facet. Since at every vertex of the icosahedron, five of its faces meet, the cut polyhedron, called a truncated icosahedron, has five times as many vertices. The truncated icosahedron therefore has 60 vertices and its faces are either identical pentagons or identical hexagons - see Fig.~\ref{truncatedIcosahedron} for an illustration. Incidentally, the truncated icosahedron is the design of the classic football and the structure of the carbon allotrope Buckminsterfullerene. The latter is often colloquially referred to as a ``Buckyball``.

	In analogy with the Platonic Bell inequalities, we obtain a Buckyball Bell inequality using the construction in Eq.~\eqref{gen}. To facilitate the fact that Alice and Bob will have 60 measurements each, we note that if a vector points to a vertex of the Buckyball, then the antipodal vector also points to a vertex of the Buckyball. Therefore, we only supply Alice and Bob with 30 measurements each, which are intended to point to the 30 vertices of the Buckyball which are not antipodal to each other. By choosing two perfectly aligned Buckyballs, the resulting Buckyball Bell inequality is
	\begin{align}
	&\mathcal{B}_\text{Buckyball} \stackrel{\text{local}}{\leq}   \frac{20}{109}\left(461+493\varphi\right)\approx 230.952 \\
	&\mathcal{B}_\text{Buckyball} \stackrel{\text{quantum}}{\leq} 300,
	\end{align}
	where the quantum bound is obtained via the hierarchy of quantum correlations and saturated by choosing the Buckyball in Eq.~\eqref{corr}. The local bound is obtained by evaluating Eq.~\eqref{cbound}.

	\section{Outperforming the CHSH Bell inequality}
	The simplest Bell inequality test requires only two measurements each for Alice and Bob. The Bell inequality which describes this setting is known as the Clauser-Horne-Shimony-Holt (CHSH) inequality \cite{CHSH}. In fact, the CHSH inequality can straightforwardly be obtained from our general form in Eq.~\eqref{gen} by choosing $\vec{v}_1=(1,0,0)$ and $\vec{v}_2=(0,0,1)$ as well as $\vec{u}_1=(1,0,1)/\sqrt{2}$ and $\vec{u}_2=(1,0,-1)/\sqrt{2}$. The CHSH inequality reads
	\begin{equation}\label{CHSH}
	\mathcal{B}_\text{CHSH}\equiv E(1,1)+E(1,2)+E(2,1)-E(2,2)\stackrel{\text{local}}{\leq} 2.
	\end{equation}
	Via Eq.~\eqref{corr}, we saturate the maximal quantum violation, $\mathcal{B}_\text{CHSH}=2\sqrt{2}$. 
	
	An interesting question is the amount of disturbance that the quantum implementation can tolerate before ceasing to violate a Bell inequality. This is commonly modeled by mixing the desired quantum state (typically, the maximally entangled state) with white noise represented by the maximally mixed state, i.e.~
	\begin{equation}\label{visi}
	\rho_v=v\ketbra{\phi^+}{\phi^+}+\frac{1-v}{4}\openone,
	\end{equation}
	where $v\in[0,1]$ is called the \textit{visibility}.  It is then relevant to find the critical visibility below which one can no longer violate a Bell inequality. In the case of the CHSH inequality, a simple computation shows that the critical visibility is $v=1/\sqrt{2}\approx 0.7071$. As it has turned out, only few Bell inequalities can outperform the CHSH inequality in terms of their critical visibility for the maximally entangled state. The first example was reported in 2008; Ref.~\cite{Tamas} constructed a Bell inequality with 465 settings on each side and showed a critical visibility of $v\approx 0.7056$. Recently, Bell inequalities with 42 settings on each side have been discovered, that further reduce the critical visibility of the maximally entangled state to $v\approx 0.7012$ \cite{Brierley}. The method for finding the latter Bell inequality relies on the development of an efficient algorithm for finding a separating hyperplane between a point and a convex set. In this context, the point is a quantum probability distribution measured in a Bell experiment and the convex set is the set of local correlations.
	
\begin{figure}
		\includegraphics[width=\columnwidth]{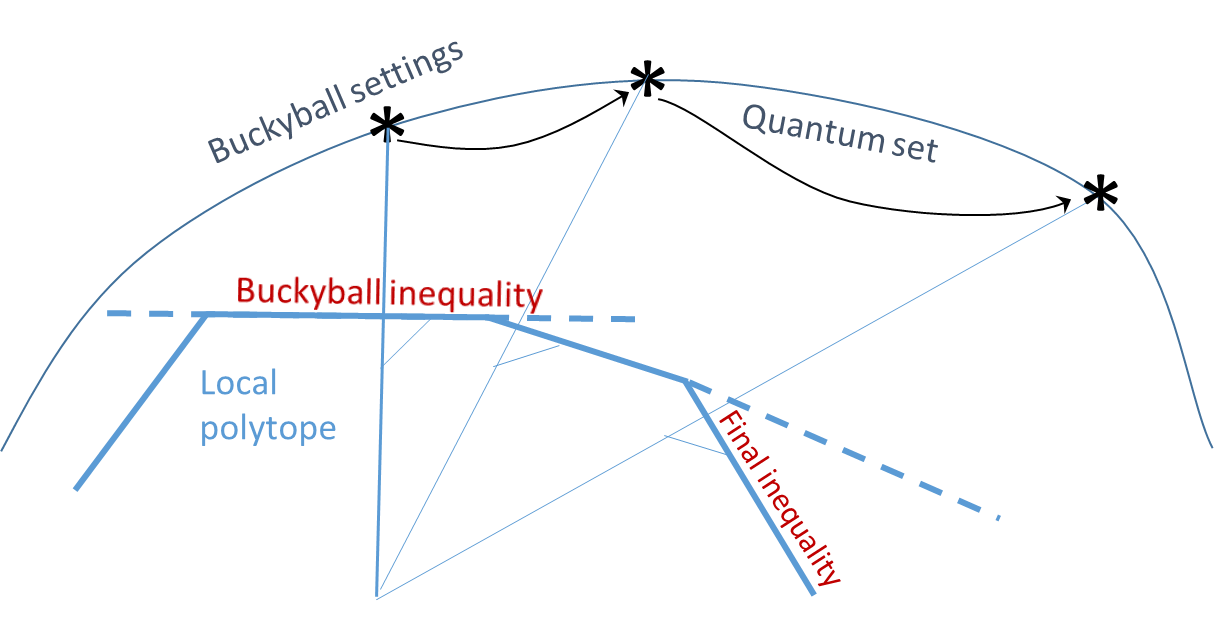}
		\caption{Illustration of our application of the algorithm of Ref.~\cite{Brierley}. Starting from the quantum probability distribution obtained from the Buckball, we find a Bell inequality that detects it. Then, we find the best quantum violation of that Bell inequality and repeat the procedure many times.}\label{algofig}
	\end{figure}

	We have implemented the algorithm of Ref.~\cite{Brierley} based on the Buckyball. Specifically, we compute the probability distribution corresponding to Alice and Bob measuring along aligned Buckyballs on the maximally entangled state. Via the algorithm, we find a hyperplane that separates it from the local set. Such a hyperplane can be written as the left-hand-side of a general Bell inequality, i.e.~as in Eq.~\eqref{Bell}. We compute the local bound associated to the hyperplane as well as the maximal quantum violation. This gives us a new probability distribution. We mix it with a small amount of noise, corresponding to Eq.~\eqref{visi}, and again run the algorithm. The procedure is repeated, and thus, noise is added and the probability distribution is perturbed, until it appears that we no longer find Bell inequalities with improved critical visibility. We illustrate the procedure in Figure~\ref{algofig}.
	Implementing this procedure based on the Buckyball, we have found a 30 setting Bell inequality with a critical visibility of $v\approx 0.7054$. Whereas we used the Buckyball as our starting point, the quantum violation that corresponds to the stated visibility is achieved with other polyhedra that have more complicated structures. Unfortunately, the Bell inequality appears not to admit a simple analytical form. However, for sake of completeness, we present it in Appendix.


	\section{Lost in beauty}
	There are different ways of reading our findings. First, there is the attractive connection established between the beautiful and historically rich Platonic solids and foundational relations in our arguably most successful physics theory, quantum mechanics. But, secondly, there is a lesson to be learned here. Mathematical beauty was our initial motivation. The derived Platonic Bell inequalities are undoubtedly very elegant. However, admittedly, they are not experimentally friendly. They require many more measurement settings than necessary and in spite of the efforts going into developing an elegant construction, their resistance to noise (which is unavoidable in any experiment) is lower than in numerous simpler Bell inequalities. Naturally, it would be nice to see the Platonic Bell inequalities be violated in  experiments; motivated simply by the appreciation of the Platonic solids and quantum nonlocality. However, unless the relevant technology incidentally happens to be set up and ready to use, it is unlikely that a practically minded experimenter would perform such an experiment. Indeed, only when we moved away from mathematical beauty, we eventually found a Bell inequality experiment (somewhat related to the Archimedean Buckyball) which is more noise resistant than the CHSH Bell inequality. The improvement is small, but it illustrates that searching to connect with experimental physics led us away from mathematical beauty. We believe that this carries a general lesson, namely that there is tension between mathematical beauty and experimentally friendly theoretical models \cite{Hossenfelder}. Mathematical beauty can  help in structuring the initial steps in new research directions, but unless theoretical models have experimental realities in mind, there is the danger of losing sight of empirical sciences.

	\begin{acknowledgements}
		We thank Flavien Hirsch for sharing his code for implementing the algorithm of Ref.~\cite{Brierley}. We thank Thors Hans Hansson, Antonio Ortu and Augustin Baas for comments on the introduction. We thank Jessica Elsa Sellin for drawing our attention to Platonic solids in the structure of silicates. This work was supported by the Swiss National Science Foundation (Starting grant DIAQ, NCCR-QSIT).
	\end{acknowledgements}

\appendix

\section{Noise-tolerant Bell inequality}
Below we give the coefficients $c_{x,y}$ for a Bell inequality of the form of Eq.~\eqref{Bell} that outperforms the CHSH inequality in terms of noise tolerance. The local bound of the Bell inequality is $  145.0181$ and a quantum violation of $ 205.5873$ is possible using a maximally entangled state. Notably, the critical visibility is the ratio of these two numbers, which is $0.7054$. We give the coefficients in two matrices: the first one covers the values $y=1,\ldots, 15$ and the latter covers the values $y=16,\ldots,30$.

\onecolumngrid
\newpage

\begin{sidewaysfigure}
	
	\begin{small}
		$\left(
		\begin{array}{c}
		\{-0.938939,0.158357,-0.533966,-0.23064,-0.724554,0.244518,-0.272051,-0.503304,-0.261808,0.424143,0.0966467,0.796627,0.188715,0.594949,-0.237468\} \\
		\{0.18072,-0.768066,-0.610507,0.60621,-0.777618,-0.237579,-0.429268,-0.464182,0.318353,-0.327486,0.807169,0.443985,0.476952,0.139788,0.647954\} \\
		\{-0.804436,-0.644061,0.152624,-0.0978691,0.195717,0.594001,0.225163,0.271193,-0.842739,0.502871,-0.743338,-0.787921,-0.416596,-0.775433,0.720186\} \\
		\{-0.265651,0.14169,-0.369637,-0.236285,0.568644,0.113412,-0.602475,0.624299,-0.226567,0.317713,-0.553164,0.54268,-0.776341,0.652845,-0.710988\} \\
		\{-0.649483,-0.470572,0.245036,0.737009,0.258046,-0.493589,0.227192,0.242118,0.305222,-0.846779,-0.61715,-0.725507,-0.799294,-0.641217,0.610627\} \\
		\{0.413947,-0.307873,0.635676,0.160502,-0.149634,-0.314452,0.51116,-0.145385,0.425781,-0.0626215,0.661908,-0.385767,0.675485,-0.781107,0.234186\} \\
		\{-0.354744,-0.683865,0.21573,-0.185906,0.294178,0.658654,0.153385,0.214849,-0.571555,0.68996,-0.43329,-0.361252,-0.230952,-0.405367,0.552939\} \\
		\{-0.568539,-0.285698,0.218428,0.575954,0.339942,-0.585974,0.15821,0.232436,0.812193,-0.43336,-0.309785,-0.250307,-0.396192,-0.213033,0.72279\} \\
		\{-0.340875,0.637664,-0.73432,-0.164406,0.34447,0.41788,-0.281528,0.567055,-0.629244,0.0903495,-0.714465,-0.799878,-0.779148,0.356293,-0.241836\} \\
		\{0.328608,-0.362466,0.291346,0.406731,-0.70639,-0.235605,0.57929,-0.512153,0.149479,-0.798215,-0.308225,-0.726373,0.641391,-0.624684,0.667658\} \\
		\{0.43759,0.89431,-0.773745,-0.548761,-0.814545,0.562296,-0.290851,-0.245683,-0.771971,-0.737208,0.259936,0.114717,0.308442,0.248691,0.232629\} \\
		\{0.787729,0.179406,-0.641468,0.691865,-0.932348,-0.605349,-0.406903,-0.297606,-0.519086,-0.67114,0.189645,0.206028,0.354553,0.731856,0.619125\} \\
		\{0.220948,0.449649,-0.667992,-0.907741,-0.716903,0.664622,-0.197348,-0.402631,-0.692509,0.462677,0.781921,0.256704,0.330691,0.127435,-0.620854\} \\
		\{0.431043,0.150873,-0.726035,0.546273,-0.349899,-0.871917,-0.296315,-0.31912,0.652939,-0.635094,0.270925,0.356846,0.208299,0.506987,0.646933\} \\
		\{-0.222491,0.597369,0.717729,-0.740314,0.494043,0.318595,0.406368,0.890863,-0.217487,0.576908,0.275939,0.731067,-0.618467,0.633616,0.0314815\} \\
		\{0.62162,-0.226652,0.678117,0.236063,0.576352,-0.728841,0.623365,0.609693,0.650236,-0.241325,0.688331,0.19918,0.600824,-0.625507,0.247508\} \\
		\{-0.629058,-0.665512,0.762582,0.699471,0.663632,0.55867,0.74883,0.594934,-0.779332,-0.666641,0.756054,0.747264,-0.335007,-0.0285627,0.696037\} \\
		\{-0.673681,0.362363,0.764493,-0.569922,0.415839,0.823724,0.424997,0.530407,-0.332561,-0.424747,0.576711,0.76404,-0.395631,0.703754,0.823685\} \\
		\{0.43872,-0.683226,0.370403,0.741479,0.803973,-0.679839,0.553606,0.61573,-0.318631,-0.37065,0.722129,0.66742,0.574685,-0.505264,0.217202\} \\
		\{-0.925456,0.542483,-0.319801,-0.719634,0.438413,0.787642,0.603134,0.868044,0.744695,0.355048,-0.147578,-0.336373,-0.54689,-0.772933,-0.427955\} \\
		\{0.661664,-0.711855,0.637209,0.599201,-0.2172,-0.921964,0.676367,0.688874,0.333304,0.476668,-0.292252,-0.267618,-0.791729,-0.213157,0.0155498\} \\
		\{-0.000473923,0.674635,-0.621596,-0.654198,-0.771299,0.222375,0.344293,0.298822,0.59341,0.489033,-0.703297,-0.355922,-0.663328,-0.612558,-0.629366\} \\
		\{0.707855,-0.439378,-0.392686,0.53223,-0.51318,-0.660753,0.753103,0.487686,0.292849,0.578796,-0.267075,-0.212718,-0.62204,-0.703102,-0.81733\} \\
		\{-0.741289,0.29658,-0.549618,-0.240212,-0.63298,0.392904,-0.298802,-0.908456,-0.382104,0.875208,0.629615,0.615326,0.697973,0.243977,-0.744801\} \\
		\{0.212004,-0.585664,-0.7033,0.32745,-0.43961,-0.258786,-0.581786,-0.370539,0.658757,-0.363462,0.263837,0.709558,0.143703,0.42078,0.58684\} \\
		\{-0.697701,0.446153,-0.667549,-0.265277,0.391646,0.394696,-0.624305,-0.659119,-0.196455,0.625685,0.820596,0.356001,0.669307,0.293896,-0.642148\} \\
		\{0.563873,-0.56564,-0.121853,0.551754,-0.588406,-0.536469,-0.697011,-0.721448,0.673423,-0.243212,0.224131,0.780953,0.176979,0.588214,0.670529\} \\
		\{0.694967,0.649393,-0.277494,-0.705304,-0.215655,-0.515298,-0.688266,-0.582478,0.66964,0.671606,0.375588,0.287088,0.806006,0.782312,-0.381033\} \\
		\{-0.222341,0.305532,0.7595,-0.795671,0.193786,0.65971,0.734857,0.403256,-0.485025,0.624732,-0.709763,-0.737372,-0.389726,-0.399119,0.209891\} \\
		\{0.509022,-0.0679171,0.259325,0.433121,0.692275,-0.867476,0.300912,0.637387,0.672749,-0.504335,-0.491735,-0.633081,-0.638331,-0.324914,0.244414\} \\
		\end{array}
		\right)$
	\end{small}
\end{sidewaysfigure}

\begin{sidewaysfigure}
	\begin{small}
		$\left(
		\begin{array}{c}
		\{0.618765,-0.662851,-0.734738,0.484761,-0.73526,0.727705,-0.567132,0.671811,-0.501899,0.219975,-0.689996,0.409004,0.678675,-0.173631,0.548544\} \\
		\{-0.205502,-0.620415,0.480766,-0.702919,0.687927,-0.760476,0.68856,0.0236075,0.204386,-0.644982,0.634582,-0.656913,0.767748,0.341393,-0.233116\} \\
		\{0.350289,0.701243,0.693346,0.330624,-0.214131,0.325917,-0.564207,-0.752727,-0.332827,-0.521661,-0.640136,0.169096,-0.150438,0.746876,0.245401\} \\
		\{0.3686,0.452042,-0.752094,0.826602,-0.802341,0.707194,-0.664093,0.211129,-0.382652,0.370586,-0.380456,0.278798,-0.467576,-0.744942,0.416623\} \\
		\{0.647496,0.664758,0.446486,0.845724,0.698577,-0.435885,-0.453555,-0.631341,-0.668105,-0.32313,-0.0426063,-0.686365,-0.252337,0.262152,0.802131\} \\
		\{-0.761503,0.639096,0.695576,-0.389685,0.582081,-0.627134,0.350879,-0.652563,0.388622,-0.162314,0.478742,-0.347407,-0.557886,0.601113,-0.929526\} \\
		\{0.866332,0.697774,0.608627,0.645319,0.624741,0.758558,0.498415,0.434939,-0.286139,-0.711873,-0.82185,-0.711271,-0.559926,0.723296,0.352421\} \\
		\{0.602424,0.770311,0.672259,0.53567,0.744653,0.7174,0.729257,0.322235,-0.728996,-0.405124,-0.635146,-0.593506,-0.529851,0.252734,0.834103\} \\
		\{0.597579,-0.681695,-0.255051,-0.409194,0.611591,0.336535,0.626865,0.303354,-0.221829,0.695675,-0.269514,0.570629,0.603758,-0.640025,0.616101\} \\
		\{-0.151168,-0.789821,-0.375987,-0.35428,0.55729,0.481758,0.313856,0.606919,0.647259,-0.238287,0.631875,-0.247064,0.634522,0.574709,-0.439511\} \\
		\{0.68519,0.620034,0.625722,0.718528,-0.212594,-0.368697,-0.329059,-0.431544,0.810066,0.406892,0.7785,0.378512,0.381378,-0.602278,-0.747872\} \\
		\{0.194688,0.768665,0.865638,0.506669,-0.467267,-0.221339,-0.38699,-0.665194,0.272847,0.560317,0.24777,0.667351,0.469105,-0.382848,-0.810405\} \\
		\{0.669664,0.00024622,-0.451122,0.69841,-0.251988,-0.679355,-0.695921,-0.609369,0.590384,0.205755,0.631482,0.302091,0.859159,-0.339175,-0.433473\} \\
		\{-0.716336,-0.264214,0.584305,-0.398949,-0.857242,-0.659392,-0.48178,-0.563143,0.236219,0.590674,0.22193,0.600616,0.85844,-0.608455,-0.22848\} \\
		\{0.22872,0.512406,0.719643,0.268033,-0.375038,-0.476979,-0.314531,-0.774318,-0.688045,0.759848,-0.676097,0.631291,-0.387913,0.622789,0.230733\} \\
		\{0.0618877,0.589273,0.211638,0.805043,-0.0651967,-0.337303,-0.728714,-0.740788,0.708116,-0.661282,0.614384,-0.682569,-0.422501,0.228226,0.261265\} \\
		\{0.601843,0.22407,0.172081,0.225597,-0.789967,-0.710092,-0.193365,-0.229389,-0.381636,-0.279267,0.700554,0.660094,-0.28382,0.285817,0.44618\} \\
		\{0.222794,0.288243,0.615222,0.172299,-0.449681,-0.787868,-0.246614,-0.145826,-0.67505,0.64223,0.467168,0.322405,-0.285842,0.686903,0.348719\} \\
		\{0.751474,0.198228,0.153561,0.712058,-0.74731,-0.315494,-0.628324,-0.128811,0.678201,-0.724899,0.381721,0.347916,-0.30515,0.269474,0.566261\} \\
		\{-0.468073,-0.794563,-0.314278,-0.776495,0.398847,0.0624419,0.307419,0.287304,-0.535865,-0.456788,-0.291028,-0.667804,0.868123,-0.611396,0.335802\} \\
		\{-0.41862,-0.830706,-0.775694,-0.469073,0.213926,0.457578,0.208123,0.597829,-0.584421,-0.39469,-0.632849,-0.15915,0.457207,0.604093,-0.620319\} \\
		\{-0.821301,-0.136446,-0.191609,-0.213651,0.476984,0.294063,0.286634,0.231716,-0.797946,-0.457097,-0.572769,-0.845424,0.753208,-0.663784,-0.665987\} \\
		\{-0.468497,-0.279621,-0.717694,-0.283827,0.275871,0.308916,0.146064,0.22691,-0.504266,-0.774309,-0.854537,-0.355399,0.811631,-0.402012,-0.626346\} \\
		\{0.98798,-0.252349,-0.713178,0.730876,-0.314607,-0.491696,-0.797988,-0.58876,-0.606883,0.137153,-0.460238,0.240625,0.794913,-0.416869,0.564035\} \\
		\{-0.684411,-0.391126,0.642472,-0.643781,-0.454216,-0.63543,-0.348454,-0.736423,0.166555,-0.586306,0.176885,-0.631294,0.573863,0.275525,-0.307261\} \\
		\{0.416442,0.603794,0.328354,0.280055,-0.215751,-0.646012,-0.327045,-0.757004,-0.704772,0.21979,-0.283136,0.156536,-0.772339,-0.64949,0.791263\} \\
		\{-0.631459,0.595693,0.356961,0.407268,-0.708888,-0.237543,-0.702033,-0.680114,0.240264,-0.408217,0.222841,-0.274835,-0.296321,0.765918,-0.74526\} \\
		\{-0.454223,-0.293342,-0.304669,-0.372679,0.479959,0.857962,0.814488,0.739922,0.706705,0.668698,-0.412643,-0.464132,0.0813487,-0.305294,-0.364783\} \\
		\{0.25035,0.300207,0.725872,0.231932,-0.67687,0.226395,-0.739863,-0.744072,-0.315295,0.477261,-0.711921,0.692551,-0.337137,0.788336,0.141295\} \\
		\{0.518217,0.476583,0.375773,0.618415,0.588361,-0.655611,-0.447799,-0.675673,0.385371,-0.353972,0.855368,-0.812284,-0.347944,0.112769,0.770458\} \\
		\end{array}
		\right)$
	\end{small}
\end{sidewaysfigure}
	
\end{document}